%% file: exec_model.tex
\documentclass[conference]{IEEEtran}

\usepackage{algorithm}
\usepackage[noend]{algpseudocode}
\usepackage{amsmath}
\usepackage{amssymb}
\usepackage{bbm}
\usepackage{booktabs}
\usepackage{color}
\usepackage{comment}
\usepackage{float}
\usepackage{graphicx}
\usepackage{ifpdf}
\usepackage[utf8]{inputenc}
\usepackage{keyval}
\usepackage[switch,columnwise]{lineno}
\usepackage{listings}
\usepackage{lscape}
\usepackage{multirow}
\usepackage{paralist}
\usepackage{rotating}
\usepackage[normalem]{ulem}
\usepackage{url}
\usepackage{xspace}
\usepackage{xcolor}
\usepackage{wrapfig}
\usepackage{algpseudocode}
\usepackage{subcaption}
\usepackage{enumitem}
\usepackage{cite}

\begin{document}

\title{Towards General Distributed Resource Selection}

 \author{Ming Tai Ha$^{1}$, Matteo Turilli$^{1}$, Andre Merzky$^{1}$ and Shantenu Jha$^{1}$$^{,2}$\\
    \small{\emph{$^{1}$ Rutgers, the State University of New Jersey, Piscataway, NJ 08854, USA}}\\
    \small{\emph{$^{2}$ Brookhaven National Laboratory}}\\
 }

\date{}
\maketitle

%----------------------------------------------------------------------------
% ABSTRACT
%----------------------------------------------------------------------------
\begin{abstract}
\input{abstract}
\end{abstract}

%----------------------------------------------------------------------------
% KEYWORDS
%----------------------------------------------------------------------------

%----------------------------------------------------------------------------
% INTRODUCTION
%----------------------------------------------------------------------------
\section{Introduction}\label{sec:intro}
\input{intro}

%----------------------------------------------------------------------------
% RELATED WORK
%----------------------------------------------------------------------------
\section{Related Work}\label{sec:rel_work}
\input{related_work}

%----------------------------------------------------------------------------
% Resource Selection
%----------------------------------------------------------------------------
\section{Resource Selection}\label{sec:selection}
\input{res_select_2}

%----------------------------------------------------------------------------
% Experiments
%----------------------------------------------------------------------------
\section{Experiments}\label{sec:experiments}
\input{experiments}

%----------------------------------------------------------------------------
% Conclusions
%----------------------------------------------------------------------------
\section{Discussions \& Conclusions}\label{sec:conclusions}
\input{conclusions}

%----------------------------------------------------------------------------
% Acknowledgment
%----------------------------------------------------------------------------
\section*{Acknowledgment}
\input{acknowledgment}

%----------------------------------------------------------------------------
% Bibliography
%----------------------------------------------------------------------------
\bibliographystyle{IEEEtran}
\bibliography{exec_model}
\end{document}

%% file: abstract.tex
The advantages of distributing workloads and utilizing multiple distributed resources are now well established. The type and degree of heterogeneity of distributed resources is increasing, and thus determining how to distribute the workloads becomes increasingly difficult, in particular with respect to the selection of suitable resources. We formulate and investigate the resource selection problem in a way that it is agnostic of specific task and resource properties, and which is generalizable to range of metrics. Specifically, we developed a model to describe the requirements of tasks and to estimate the cost of running that task on an arbitrary resource using baseline measurements from a reference machine. We integrated our cost model with the Condor matchmaking algorithm to enable resource selection. Experimental validation of our model shows that it provides execution time estimates with 157--171\% error on XSEDE resources and 18--31\% on OSG resources. We use the task execution cost model to select resources for a bag-of-tasks of up to 1024 GROMACS MD simulations across the target resources. Experiments show that using the model's estimates reduces the workload's time-to-completion up to \(\sim\)85\% when compared to the random distribution of workload across the same resources.

%% file: intro.tex
The Worldwide Large Hadron Collider Grid (WLCG) was created to support
multiple experiments at LHC that collect and distribute hundreds of petabytes
of data worldwide to hundreds of computer centers. The WLCG has become one of
the prototypical example of utilizing distributed and heterogeneous resources
for scientific computing. There are many other large scale experimental and
observation facilities---SKA, LSST, DUNE to name just a few---that also
require multiple heterogeneous resources, although the number and type of
resources varies.

The workloads from these experiments are often expressed as a collection of
tasks. Historically, these experiments have placed tasks on different
resources using implicit assumptions about resources and task properties. The
types of tasks and distributed resources that need to be federated is
changing; however, existing customized algorithms and approaches are not
easily generalizable, nor extensible.

The availability of multiple distributed resources, often with diverse
capabilities, offers the opportunity to improve resource utilization and
increase the concurrency of workload execution. The benefits of distributing
the execution of a workload across several types of resources has been
investigated~\cite{calatrava2011combining,deelman2010grids}, however, a
consequence of the increase in the number and types of distributed resources
is the resource selection problem. This problem can be formulated as: ``The
selection of a subset of resources to execute a workload among those
available to a user''. Resource selection is composed of two questions: one
question is of resource viability, which asks ``Which resources can be used
to execute the given workload?''; the other is of execution affinity, which
asks ``Which available resources should be used in the execution of the
workload?''.

The resource viability problem has been addressed by~\cite{match1998}, which
provided a general method that uses the requirements of tasks and
capabilities of a resource to determine whether a resource can successfully
execute the task. The execution affinity problem has been investigated using
application benchmarks comprised of a pre-defined suite of applications to
provide an understanding on how resources perform~\cite{benchmark2014}.
However, further work is required to standardize the process by which
performance data is measured, analyzed, reported and
interpreted~\cite{benchmark2014}. Workload scheduling
algorithms~\cite{zhou2016monetary,rodriguez2017scheduling} either implicitly
need to solve the execution affinity problem, or require it to be solved.
These algorithms need knowledge of the cost of executing each task on each
viable resource. Approaches to estimate task execution costs either require
information on the task's code structure and hardware architecture, or
historical data on the cost of running each task on the possible target
resources.

Several studies compare different types of applications and resources, and
how different applications can exploit different types of
resources~\cite{juve2013comparing,hwang2016role,hussain2013hpdcsurvey}.
However, there are no general models for the resource selection problem, or
published results benchmarking them compared to a random selection of
resources. Thus, not surprisingly, there do not exist quantitative estimates
of improvements arising from resource selection as a function of scale
(number of tasks and resources), or the degree of heterogeneity. Last but not
least, resource selection methods have been tightly integrated with specific
software tools, making comparisons across resource selection algorithms
intractable.

Against this backdrop and its increasing importance, we investigate the
generalized distributed resource selection problem. We focus on a class of
problems that have not historically utilized distributed resources: A large
fraction of the approximately half a million single core jobs submitted to
XSEDE are MD simulations using community codes such as GROMACS, AMBER and
CHARMM\@. These simulations are typically bound to a specific resource at the
time of job submission. Assuming resources are fungible, the suitable
selection of resources raises the theoretical possibility of reducing the
time-to-completion of the tasks and improving the overall throughput of the
collective set of XSEDE resources. This requires selecting resources that
allow to take advantage of lower queue waiting times while not incurring into
a higher execution waiting time, or vice versa. Both scenarios reiterate the
importance of resource selection for the collective utilization of
distributed computing resources.

In this paper, we formulate and investigate the resource selection problem
that is agnostic of specific task and resource properties, and which is
generalizable to a range of metrics (such as monetary cost, execution time
etc.). Specifically, we propose a task execution cost model which can predict
the cost of executing a task on a resource. We incorporate this model into
the Condor matchmaking process to distribute workloads across resources.
Using our model, we use historical information of the execution times of a
singly-threaded GROMACS MD simulation running on a baseline cluster (Amarel)
to predict the execution times of the same simulation on the different
resources: XSEDE (Bridges, Comet, SuperMIC) and the XSEDE OSG virtual
organization (or OSG in short). Our experiments show that our model provides
execution time estimates with 157--171\% error on XSEDE resources and
18--31\% on OSG when we do not consider any information about the
instruction-level parallelism the simulation exploits during execution. We
use the task execution cost model to select resources for a bag-of-tasks of
up to 1024 GROMACS MD simulations across the target resources. Experiments
show that using the model's estimates reduces the workload's
time-to-completion up to \(\sim\)85\% when compared to the random
distribution of workload across the same resources.

%% file: related_work.tex
The resource selection problem has been actively studied by the HTC
community. Due to the heterogeneity and transience of the resources composing
an HTC DCR, the processed of resource selection used by the HTC community
primarily focuses on: (i) standardizing how jobs and resources are
described~\cite{jsdl-spec,classad-spec,wsrf-spec,rsl-spec,glue2-spec}; (ii)
providing ways to discover the resources available in a HTC
DCR~\cite{mds2001,nws1999}; and (iii) providing a general method, known as
matchmaking, to match jobs with resources that satisfy their
requirements~\cite{match2001}. Often, the process of matchmaking is carried
out by resource brokers~\cite{condorg2001,elmroth2004benchbroker,kim2004design,nimrodg2000}.

Application benchmarking is most often related to the problem of estimating
the performance of an application workload on a given resource of group of
resources. Application benchmarks provide insights into how a specific
resource performs when executing a predefined application workload. The use
of application
benchmarks~\cite{elmroth2004benchbroker,clematis2010benchbroker} to express
the requirements of a workload has been shown to improve the performance of
the matchmaking process. However, given the state-of-the-art, much effort is
still required to standardize how to measure, analyze and report benchmark
results~\cite{benchmark2014}. Without such a standardization, comparison and
interpretation of benchmarks across multiple resources remains difficult.

Work has been done on providing bounds or estimates on the worst-case
execution times of an application workload on a given
resource~\cite{wilhelm2008worst,abella2014comparison,abella2015wcet}. Given a
workload with a set of tasks, deterministic methods for timing analysis
analytically derive an upper bound on the execution time of each task, based
on hardware specifications and tasks' code structure~\cite{wilhelm2008worst}.
Probabilistic timing analysis use both static and measurement-based methods,
executing workload's tasks on the hardware of a resource with a predefined
set of inputs to estimate the worst-case execution time of the task.

Deterministic and probabilistic timing analyses suffer from limitations
related to the amount and accuracy of the information they require, or to the
strength of their assumptions. Deterministic methods require detailed
knowledge of the resource's hardware and of the task's code structure;
probabilistic analyses using static methods assume information on the task's
code structure~\cite{wilhelm2008worst}; and probabilistic analyses using
measurement-based methods do not assume any information on software
internals, but place strong assumptions on the hardware and on how
observations are taken~\cite{abella2014comparison,cazorla2013upper}.
Collecting the required information or matching the assumptions made by these
methods is challenging if not infeasible when considering selections over
multiple production-grade resources.

There have also been efforts to predict task execution time on a resource
using least-squares~\cite{kishimoto2004execution}, 
k-Nearest Neighbors~\cite{iverson1999statistical}, 
iverson1996runneural networks~\cite{duan2009hybrid}
maximum likelihood estimation and random
forests~\cite{chirkin2017execution}. These methods require historical
information of similar runs to generate predictions, and are considered a
convenient way to estimate the cost of executing a task on a resource.
However, there is limited availability of extensive and consistent
collections of historical data about the execution costs of workloads running
on production-grade resources. This reduces the usefulness of these methods
for real-life use-cases, especially when considering resource selection among
multiple resources.

%% file: res_select_2.tex
We address the resource selection problem by devising a model to predict the
cost of executing a given task on a given resource. We use this prediction,
alongside other type of information when available, to choose resources that
are more likely to optimize a given metric. In this way, we frame the
resource selection problem as follows: Given a set of tasks, a set of
resources, and a metric of performance, the resource selection problem
consists in selecting the resources that optimize the given metric when
executing the given tasks.

The predictions of our model use baseline measurements on a single reference
resource and require no information about the structure of the code executed
on that resource. We execute each task of a workload application on a
resource, collecting data about its performance. On the base of these data,
we analytically predict the cost of executing those tasks on an arbitrary
resource. In this way, we avoid to execute and measure the performance of the
same set of tasks on multiple resources.

We predict the cost of executing a task on a resource without using
information about the structure of the task's code and without detailed
information about the resource's hardware architecture. In this way, we trade
off between simplifying the process of collecting the data required to
predict the cost of execution and the accuracy of that prediction. Lack of
accuracy is acceptable as far as the model's predictions support effective
resource selection for the distributed execution of a given workload over a
given set of resources.

% ---------------------------------------------------------------------------
\subsection{Cost Model of Task Execution}\label{sec:cost_model}

The key idea of our cost model of task execution is to explicitly define the
functionalities which a task uses during its execution. We define a
consumable to be the entity that represents a functionality from a resource
which a task uses during its execution. We show that we can calculate the
cost of executing a task on a resource based on the consumables the task uses
to run to completion, and on the cost of using each consumable on a resource.

We formally define task, workload, and resource in terms of consumables, and
provide a model to estimate the cost of executing a task. Cost evaluation can
be based on multiple units of measurements such as quantity of allocation,
currency, or energy. To simplify the construction of our model, we assume:
(1) tasks use required consumables independent from the resource that offer
them; and (2) the cost of using any consumable offered by a resource is
fixed. Here the terms we define from first principles:

\begin{description}[align=left]
	\item [Consumable:] An entity representing a unit of work. A consumable
	has two properties: (1) type, which determines the kind of work the
	entity can perform; and (2) form, which specifies the conditions that
	must be satisfied for the consumable to be used to perform work.
	\item [Requirement:] Amount of a consumable, where the amount is assumed
	to be fixed.
	\item [Instruction:] Set of tuples, each specifying a certain
	requirement.
	\item [Task:] Sequence of instructions, executed in the order specified
	by the sequence.
	\item [Workload:] Set of tasks, where all tasks can run concurrently.
	\item [Capability:] Rate at which a consumable is offered, assumed to be
	fixed.
	\item [Resource:] Set of tuples, each specifying a certain capability.
\end{description}

Formally, a consumable \(c\) is a set \( \{type, form\} \), where \(type\) is
a single value while \(form\) is a set of pairs. For any pair in \(form\),
the first element is an attribute \(attr\) that uniquely identifies the pair
in \(form\); the second element of the pair is a condition \(cond\),
expressed as a set of values, that specifies how \(c\) can be used. A
requirement \(req\) is a tuple \( (c, amt)\) where \(c\) is a consumable and
\(amt > 0\) is fixed amount of that consumable. An instruction \(ins\) is a
set  \( \{req_1, \ldots, req_n\} \), where each requirement \(req_n\)
specifies the amount of a consumable required by \(ins\). A task \(task\) is
a sequence of instructions \([ins_1, \ldots , ins_n]\), where \(|ins_i|\) is
the number of requirements of each instruction
\(ins_i\). 

Let \(C\) be the set of \(m\) distinct consumables, where each consumable is
specified by a requirement of an instruction in \(task\). For each \(c \in
C\), we calculate the total amount \(amt_c\) of consumable \(c\) required by
the instructions in \(task\) by taking the sum of the amounts specified by
any requirement of any instruction that also specifies \(c\) as its
consumable. We express \(amt_c\) as:

\begin{equation}\label{eq:aggr_proc} 
amt_c = \sum_{i=1}^{n}\sum_{j=1}^{|ins_i|}
ins_i.ireq_j.amt \cdot 
\mathbbm{1}_{\{\substack{ins_i.ireq_j.c = c}\}} , 
\end{equation}

where \(\mathbbm{1}\) is the indicator function, which equals 1 if the
consumable specified by the \(j\)-th requirement of the \(i\)-th instruction
of \(task\) is \(c\), and 0 otherwise. We call Eq.~\ref{eq:aggr_proc} the
aggregation procedure.

We can now define task \(task\) also as a set \( \{req_1, \ldots, req_n\}\),
where each requirement \(req_n\) specifies the total amount of consumable \(c
\in C\) required by the instructions of \(task\). It is important to note
that the aggregation procedure discards information on the order (and
concurrency) with which the task's instructions can use consumables, but
provides a simple representation of a task's requirements. We define a
workload \(WL\) as a set of tasks \( \{task_1, \ldots, task_n\}\).

We define a capability \(cap\) as a set \(\{c, rate\}\), where \(c\) is a
consumable and \(rate > 0\) is fixed. \(rate\) represents the number of
consumables offered per unit of cost (e.g., in terms of time, money, or
energy). In this way we establish a relationship between the use of a
consumable and the cost of using a consumable. We define a resource \(res\)
as a set \(\{cap_1, \ldots, cap_n\}\), where each capability \(cap_n\) offers
the use of a unique consumable at a fixed rate.

Assuming that a given task can run on a given resource, we define the cost
\(K\) of running a task on a resource as the total cost required to
sequentially consume the amounts of consumables specified by the task
requirements, at the rate offered by the corresponding resource capabilities.
The cost of running the task on a resource is expressed sequentially because
our model does not consider the order (or concurrency) with which tasks can
use consumables, nor our model takes into account the order in which the
resource offers consumables. 

Formally, let there be a task \(task = (req_1, \ldots, req_n)\) and resource
\(res = (cap_1, \ldots, cap_n)\). Then, \(K\) is defined as:

\begin{equation}
\label{eq:cost_func} K= \sum_{i=1}^{m}\sum_{j=1}^{n}
\frac{req_i.amt}{cap_j.rate}
\mathbbm{1}_{\{\substack{req_i.c = cap_j.c}\}}
\end{equation}

% ---------------------------------------------------------------------------
\subsection{Resource Selection Process}

We decompose the problem of resource selection in two subproblems: (1)
selecting viable resources to execute a given workload; and (2) selecting a
subset of these resources that can execute the given workload, optimizing a
given metric of performance. We first adapt the Condor matchmaking algorithm
to operate in terms of consumables; we call the resulting algorithm the
`adapted matchmaking algorithm' (AMA). We then show that it is possible, but
not necessary, for the AMA to select resources based on requirements and
capabilities.

% ------------
\subsubsection{Resource Viability}

To adapt the Condor matchmaking algorithm to operate in terms of consumables,
we define the algorithms \(SATISFY\_REQ\) and \(SATISFY\_TASK\) to determine
whether a resource can execute a task. \(SATISFY\_REQ\)
(Alg.~\ref{alg:satisfy_req}) takes as input a requirement and capability and
determines whether the consumable specified by the capability can be used to
satisfy the requirement. \(SATISFY\_REQ\) checks whether: (1) the types of
consumables of the requirement and capability are the same; (2) for every
form attribute in the requirement there is a corresponding form attribute in
the capability; and (3) for each form attribute that is in both the
requirement and the capability there is a form condition in the capability
that is also a form condition in the requirement. Note that the comparison
operator used to decide whether two values are equal depends on the data type
of the values (e.g., integer, float, string), as discussed in several
specifications~\cite{jsdl-spec,classad-spec}.

\begin{algorithm}
\caption{Check if capability (\textit{cap}) consumable can be used to satisfy
requirement (\textit{req})}\label{alg:satisfy_req}
	\begin{algorithmic}[1]
		\Require \textit{req}; \textit{cap}
		\Ensure \textbf{True} or \textbf{False}

		\Procedure{satisfy\_req}{\textit{req}, \textit{cap}}
		\If {\textit{req.c.type} != \textit{cap.c.type}}
		    \State return \textbf{False}
		\EndIf
		
		\ForAll{(\textit{attr}, \textit{cond}) \textit{in} \textit{req.c.form}}
		        \If {\textit{attr} not in \textit{cap.c.form}}
		            \State return \textbf{False}
		        \EndIf
		        \If {\textit{cap.c.form.cond} \(\cap\) \textit{cond} \(\emptyset\)}
		            \State return \textbf{False}
		        \EndIf
		\EndFor
		\State return \textbf{True}
		\EndProcedure
	\end{algorithmic}
\end{algorithm}

\(SATISFY\_TASK\) (Alg.~\ref{alg:satisfy_task}) takes as input a
task and resource and checks whether for each requirement of the task there
is a capability of the resource that can satisfy the given requirement.
\(SATISFY\_TASK\) uses \(SATISFY\_REQ\) to determine whether a capability can
be used to satisfy a requirement. If \(SATISFY\_TASK\) returns \(True\) for a
given task and resource, then the task can execute on that resource.

\begin{algorithm}
\caption{Check whether a task (\(task\)) can execute on a resource (\(res\))}\label{alg:satisfy_task}
	\begin{algorithmic}[1]
		\Require \(task=(cap_1, \ldots, cap_m)\); \(res=(req_1, \ldots, req_n)\)
		%; \(m, n \in \mathbb{N}\)
		\Ensure \textbf{True} or \textbf{False}

		\Procedure{satisfy\_task}{\textit{res}, \textit{task}}
		\ForAll{{\textit{req} \text{in} \textit{task}}}:
		        \State match $\leftarrow$ $0$
		        \ForAll {\textit{cap} \text{in} \textit{res}}
		                \If {SATISFY\_REQ(\textit{req}, \textit{cap}) $=$ \textbf{True}}
		                    	\State match $\leftarrow$ $1$
		                \EndIf
		        \EndFor
		        \If {match $= 0$}
		                \State return \textbf{False}
		        \EndIf		        
		\EndFor
		\State return \textbf{True}
		\EndProcedure
	\end{algorithmic}
\end{algorithm}

Given a workload and a set of resources, \(SATISFY\_TASK\) can determine for
each task of the workload whether there is a subset of resources that can
execute that task. We call this subset of resources the ``viable resources
set'' of that task. If every task in the workload has a nonempty viable
resources set, then the workload can be executed across a subset of the
available resources.

% ------------
\subsubsection{Execution Affinity}

We assume a workload, a set \(vrs\) of viable resources \(\{res_1,
\ldots, res_n\}\) for each task \(task_n\) of that workload, and a function
\(affinity\) which maps a set of input tuples to a set of values (e.g.,
\(\mathbb{R}\)): The higher the value, the better \(res_n\) is for
executing \(task_m\). Note that for every pair of \((res_n,task_m)\), there
is only one input tuple that is used to determine the affinity value of that
\((res_n,task_m)\) pair.

Generally, the input of \(affinity\) does not necessarily include task
requirements or a resource capabilities. However, if we want to select
resources using only information about task requirements and resource
capabilities, we can use Eq.~\ref{eq:cost_func} to select resources and
provide the task requirements and resource capabilities as input to
\(affinity\).

\(RES\_SELECT\) (Alg.~\ref{alg:res_select}) identifies the resource(s) of a
set of task's viable resources that gives that highest affinity value. We
define the set \(vrs\_id = {res\_id_1, \ldots, res\_id_n}\) to be the set of
unique IDs of every resource in a task's viable resources set. We also define
the task input set \(TIS = {input_1, \ldots, input_n}\), where
\(input_i\) is the input tuple to \(affinity\) associated with \(res\_id_i\).
\(RES\_SELECT\) takes as input \(vrs\_id\), \(TIS\) and \(affinity\), and 
returns \(res \in vrs\_id\), whose associated input tuple gives the highest 
affinity value. 

\begin{algorithm}
\caption{Determines the resources on which a task input set (\(TIS\)) should
execute, given the viable set (\(vrs\_id\)) of each task of
\(TIS\)}\label{alg:res_select}
	\begin{algorithmic}[1]
		\Require \begin{flushleft} $vrs\_id = \{res\_id_1, \ldots,
		res\_id_n\}$; $TIS = \{input_1, \ldots, input_n\}$;
		$\textit{affinity}()$ \end{flushleft}
		\Ensure Resource ID $\textit{res}$

		\Procedure{res\_select}{\textit{res\_id}, \textit{TIS}, \textit{affinity}}
		\State best\_res $\leftarrow$ \textbf{NONE}
		\State best\_select\_val $\leftarrow$ $-\infty$
		\For {i from 1 to n}
		        \State select\_val = \textit{affinity(\(input_i\))}
		        \If {best\_select\_val \(<\) select\_val}
		                \State best\_select\_val \(\leftarrow\) select\_val
		                \State best\_res $\leftarrow$ \(res\_id_i\)
		        \EndIf
		\EndFor
		\State return best\_res
		\EndProcedure
	\end{algorithmic}
\end{algorithm}

\(RES\_SELECT\) can be used in conjunction with \(SATISFY\_TASK\) to
determine for every task in a workload: (1) Whether there is a nonempty
viable resources set for the given task; (2) Which resource in the viable
resources set yields the highest affinity value.

We assume that the user is able to acquire enough resources to execute each
task on their best resources, and that each task can execute independently
from each other. Since we are also only investigating how to perform resource
selection on workloads, we assume that each task runs on the resource that
yields the highest affinity value. It should be noted that the problem of
resource selection is different from the problem of scheduling tasks on the
selected resources. There is a large body of literature on task and workload
scheduling~\cite{zhou2016monetary,rodriguez2017scheduling}; a discussion in
this direction is beyond the scope of the paper.

%% file: experiments.tex
We perform two sets of experiments to characterize the accuracy of the model
we introduced in Sec.~\ref{sec:selection}. The first set characterizes the error
of our model when predicting the cost of executing a task on diverse DCRs. We
express the cost as the time taken by the task to execute, indicated by
\(T_x\). The second set of experiments compares the cost of operating
resource selection on the basis of our model's predictions to the cost of a
random resource selection. We use time-to-completion \(TTC\) of a workload
distributed across multiple and heterogeneous DCRs as measure of cost,
determined as a function of the \(T_x\) of all the workload's tasks.

% ---------------------------------------------------------------------------
\subsection{Characterization of \(T_x\)}\label{sec:tx_char}

We designed a set of experiments to characterize the accuracy of our model in
predicting the execution time \(T_x\) of a task on the XSEDE OSG
VirtualCluster resource pool (hereafter just OSG), XSEDE HPC machines, and
the Rutgers Amarel cluster. In the experiments, we used a task simulating the
dynamics of a protein in water (i.e., MD simulation), a task routinely
executed on diverse types of machines. We used GROMACS 5.0, compiled with
single-precision floating-point and SSE4.1 SIMD instructions. Though there
are newer versions of GROMACS, this is the version supported by OSG, where we
have limited or no control over the software environment. Further, since OSG
is primarily designed for loosely-coupled, single-threaded jobs, we executed
GROMACS simulations with a single thread and a single process on all DCRs.

We used the Amarel cluster as baseline machine, collecting information to
predict the \(T_x\) of a task. Amarel offered rapid access to its resources
but we could have used any other machine as baseline. We executed the same
task on three XSEDE HPC machines (Bridges, Comet, SuperMIC) and on OSG to
test the accuracy of our model's \(T_x\) predictions. For our experiments, we
used the compute nodes of Amarel, and submitted jobs to the \textit{RM},
\textit{compute}, and \textit{workq} queues of, respectively, Bridges, Comet
and SuperMIC\@. Though OSG is a heterogeneous collection of machines, we use
the term `target machines' to mean the XSEDE HPC machines and OSG resource
pool.

In our experiments, we focused on computational requirements as GROMACS is a
`compute-heavy' task with limited I/O load in the configuration we used.
Accordingly, we defined a compute-type consumable, i.e., a cycle that can
only be consumed on CPUs that support the x86 instruction set (x86 in short).
According to the definition given in Sec.~\ref{sec:selection}, we defined the
task of our experiments as a set of one requirement, where the requirement
specifies a fixed amount of cycles that need to be consumed on CPUs
supporting x86. Similarly, we defined a resource to be a set with one
capability representing the clock speed of a CPU that supports x86.

% ------------
\subsubsection{Experimental Setup}

We executed the same MD simulation for 1000, 5000, 10000, 25000, 50000,
75000, and 100000 timesteps. Each MD simulation executed on a node of Amarel
and was profiled with \texttt{perf}~\cite{perf}. We repeated each simulation
between 35 and 60 times for each number of timesteps, profiling the number of
instructions, cycles, and instructions per cycle (i.e., instruction rate). We
also profiled the average clock speed measured during the simulation's
execution and measured the simulation execution time (\(T_x\)). We used
\texttt{perf} to measure the task's execution time \(T_x\) on the XSEDE HPC
machines, but on OSG we used the wall time measurements in the log files of
the GROMACS simulations. This is because we have little control over the
software environment of OSG resources, and only \(\sim\)1.2\% of
\(\sim\)11000 trial runs were able to run both \texttt{perf} and GROMACS\@.

We used the number of cycles and the instruction rate of the MD simulations
profiled on Amarel to predict the number of cycles needed to execute those
simulation completely sequentially (i.e., no instruction-level parallelism).
We then used this prediction and information about CPU clock speed to predict
the \(T_x\) of MD simulations (i.e., tasks) when executing on the target
machines. We compared the number of instructions and cycles used when
executing the task on the target machines with those measured when executing
the task on Amarel. In this way, we measured the delta between the actual
number of cycles used by a task the number of cycles we predicted.

The \(T_x\) of a task varies depending on the clock frequency at which the 
resource's processor operates when executing the task. Accordingly, we 
predict the \(T_x\) of a task using the base and maximum clock frequencies of
the processor of the resource. We denoted these values \(base\) and \(max\),
and we used \(T_{x, base}\) and \(T_{x, max}\) to denote predictions of 
\(T_x\) made using \(base\) and \(max\), respectively.

We used XSEDE documentation and processor specifications to identify 
\(base\) and \(max\) of the processors of the XSEDE HPC machines. Since OSG
is a pool of heterogeneous resources, we represented \(base\) and \(max\) as
the weighted averages of the base and maximum clock frequencies of the
processors offer by the OSG resources. To calculate the weighted averages, we
collected information on the processors available in the OSG resource pool at
the beginning of the experimental campaign. We denote the average clock
frequency measured when executing a task on a resource as \(avg\).
Table~\ref{tab:clkspeed} shows the values for \(base\), \(max\), \(avg\) and
the sample standard deviation of \(avg\) (given in parentheses), for the
target machines.

\begin{table}
\scriptsize
\centering
\caption{Clock Speeds, in GHz}\label{tab:clkspeed}
\begin{tabular}{cccccc}
\toprule
DCR       & \(base\) & \(max\) & \(avg\)       \\
\midrule
Bridges   & 2.30     & 3.30    & 2.732 (0.038) \\
Comet     & 2.50     & 3.30    & 2.888 (0.001) \\
SuperMIC  & 2.80     & 3.60    & 3.589 (0.002) \\
OSG       & 2.50     & 3.09    & 2.930 (0.227) \\
\bottomrule
\end{tabular}
\end{table}

% ------------
\subsubsection{Equations}

We account for differences between the predicted and actual execution time
(\(T_x\)) of a task by showing that the error in our predictions is due to
the instruction-level parallelism exploited by the task.

We define \(\#instr\) as the number of instructions the task executes. Since
the only requirement of the task is that it consumes some amount of cycles,
we define:

\begin{equation}\label{eq:instr}
    \#instr = \#cycles \times instr\_rate,
\end{equation}

where \(\#cycles, instr\_rate\) denote the number of cycles used to execute
the instructions and the average number of instructions executing per cycle,
respectively. When only one instruction uses a cycle at any point in time,
\(instr\_rate = 1\).

We define \(pred\_\#cycles, act\_\#cycles\) as the predicted and actual
number of cycles used, respectively. Similarly, we define
\(instr\_rate\_pred, instr\_rate\_act\) as the predicted and actual number of
instructions executed during the period of a cycle. Since we are comparing
the execution of the same task, \(\#instr\) is fixed. 

From Eq.~\ref{eq:instr}:

\begin{equation}\label{eq:pred_to_act}
    \frac{instr\_rate\_act}{instr\_rate\_pred} = \frac{pred\_\#cycles}{act\_\#cycles},
\end{equation}

We define \(p2a\_cy = \frac{pred\_\#cycles}{act\_\#cycles}\) to be the ratio
between the predicted and actual number of cycles used. Since our model
assumes that only one instruction uses a cycle, \(instr\_rate = 1\):

\begin{equation}\label{eq:p2a_instr_rate}
    p2a\_cy = instr\_rate\_act
\end{equation}

We use Eq.~\ref{eq:p2a_instr_rate} to derive the number of cycles necessary
to execute a task sequentially on any resource. We define \(\varepsilon\) as
the percent error between \(p2a\_cy\) and \(instr\_rate\_act\) to measure how
much instruction-level parallelism affects the model's overprediction:

\begin{equation}\label{eq:p2a_pcterr}
    \varepsilon = \frac{|p2a\_cy - instr\_rate\_act|}{p2a\_cy} \times 100
\end{equation}

If the overprediction in the number of cycles required is completely due to 
the instruction-level parallelism, then \(\varepsilon = 0\).

% ------------
\subsubsection{Experimental Results}

We find that for any number of experimental timesteps, the number of
instructions required to execute a GROMACS simulation on resources from the
XSEDE HPC machine is within \(\sim\)3\% of that required when using Amarel.
However, the number of instructions required to execute a GROMACS simulation
using resources from OSG is on average 22--24\% more than that when using
Amarel. As such, we analyze data from XSEDE HPC DCRs and OSG separately.

\begin{figure}
    \centering
    \includegraphics[width=0.47\textwidth]{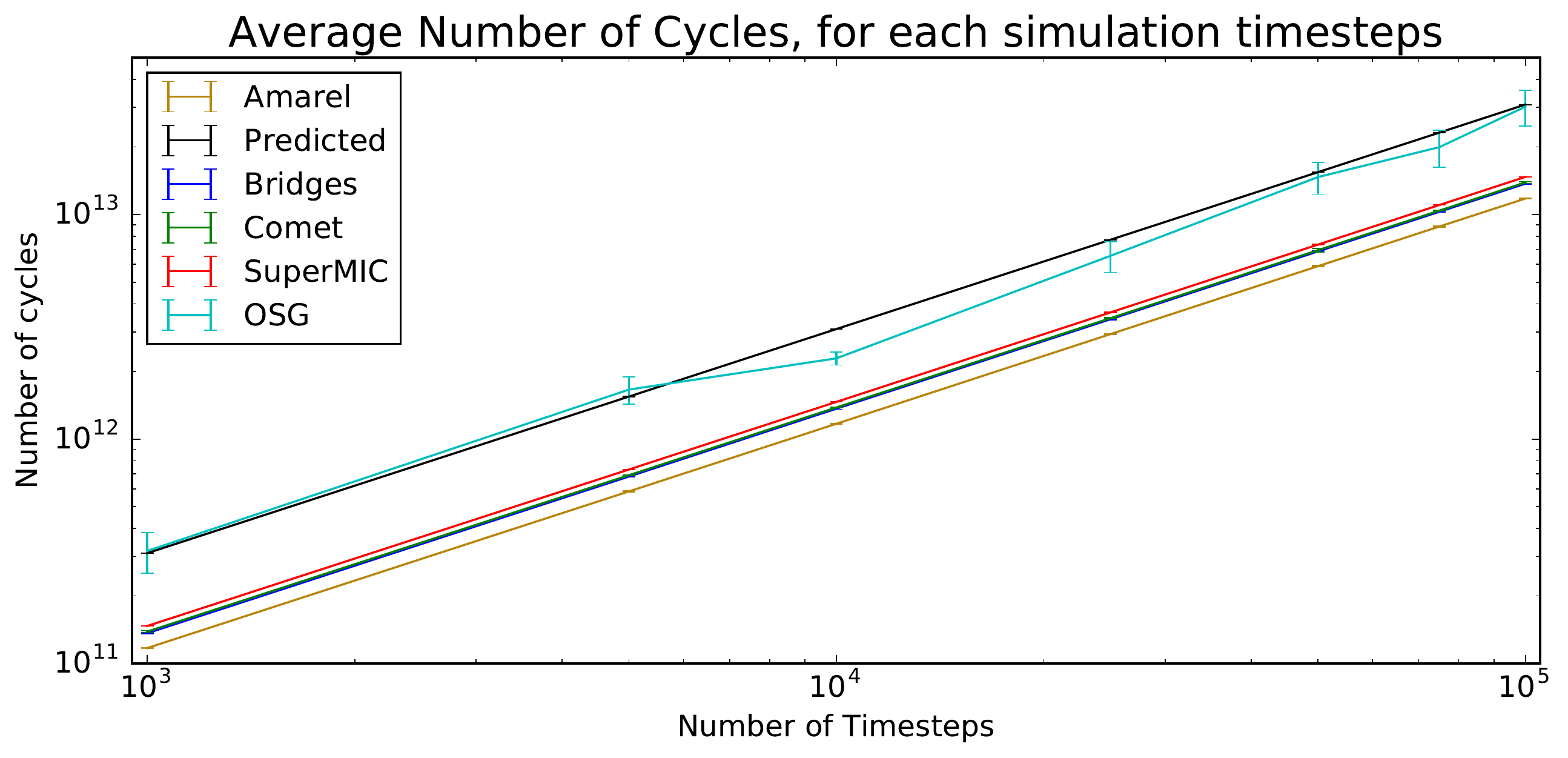}
    \caption{Average number of cycles measured on Amarel, Bridges, Comet, 
    SuperMIC, OSG, as well as the average predicted number of
    cycles}\label{fig:avg_cycles}
\end{figure}

\begin{figure}
    \centering
    \includegraphics[width=0.47\textwidth]{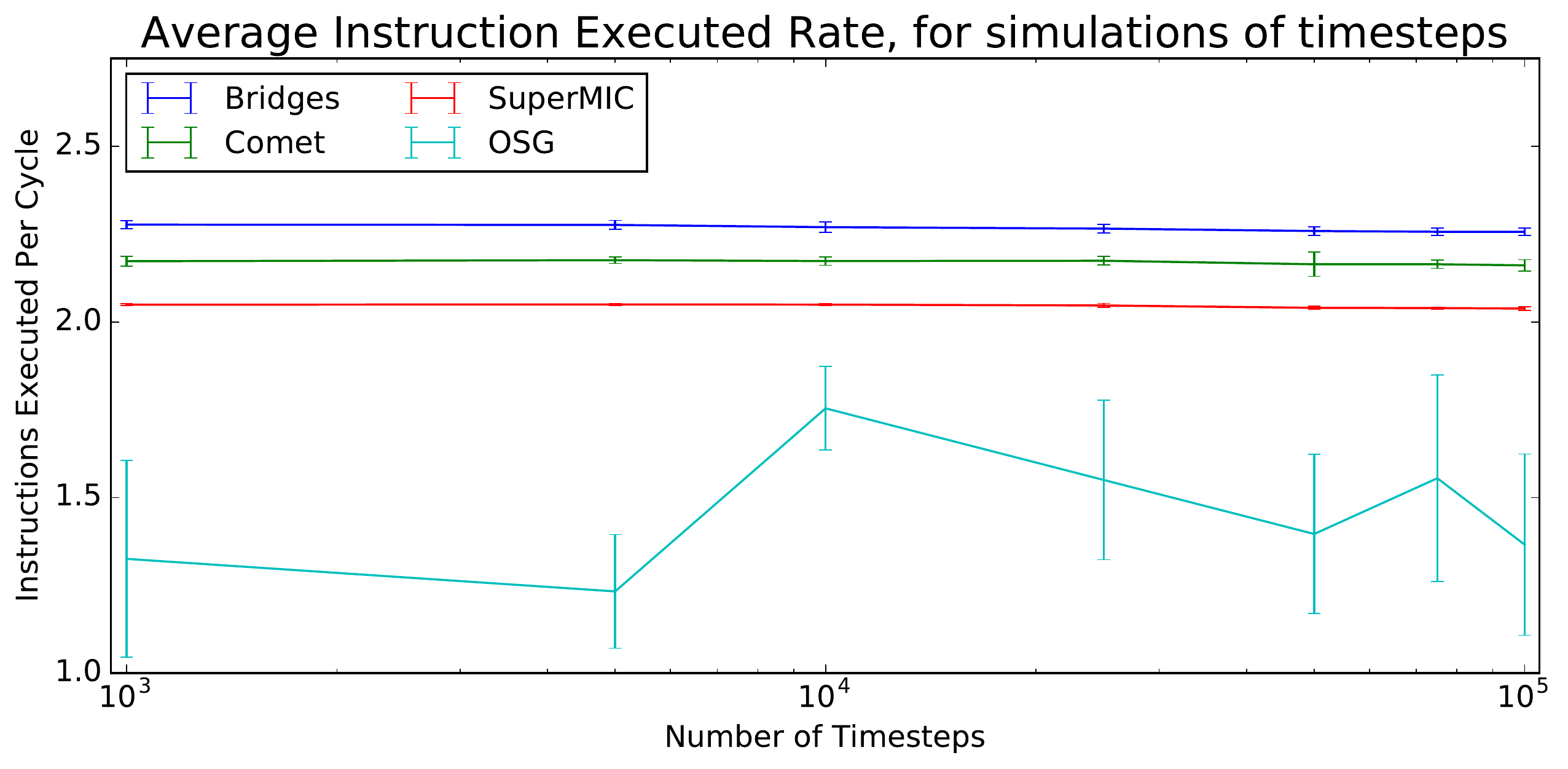}
    \caption{Instruction Rate (Instructions executed per cycle) of
    GROMACS simulations running on Bridges, Comet, SuperMIC and
    OSG}\label{fig:avg_instr_rate}
\end{figure}

\begin{figure}
    \centering
    \includegraphics[width=0.47\textwidth]{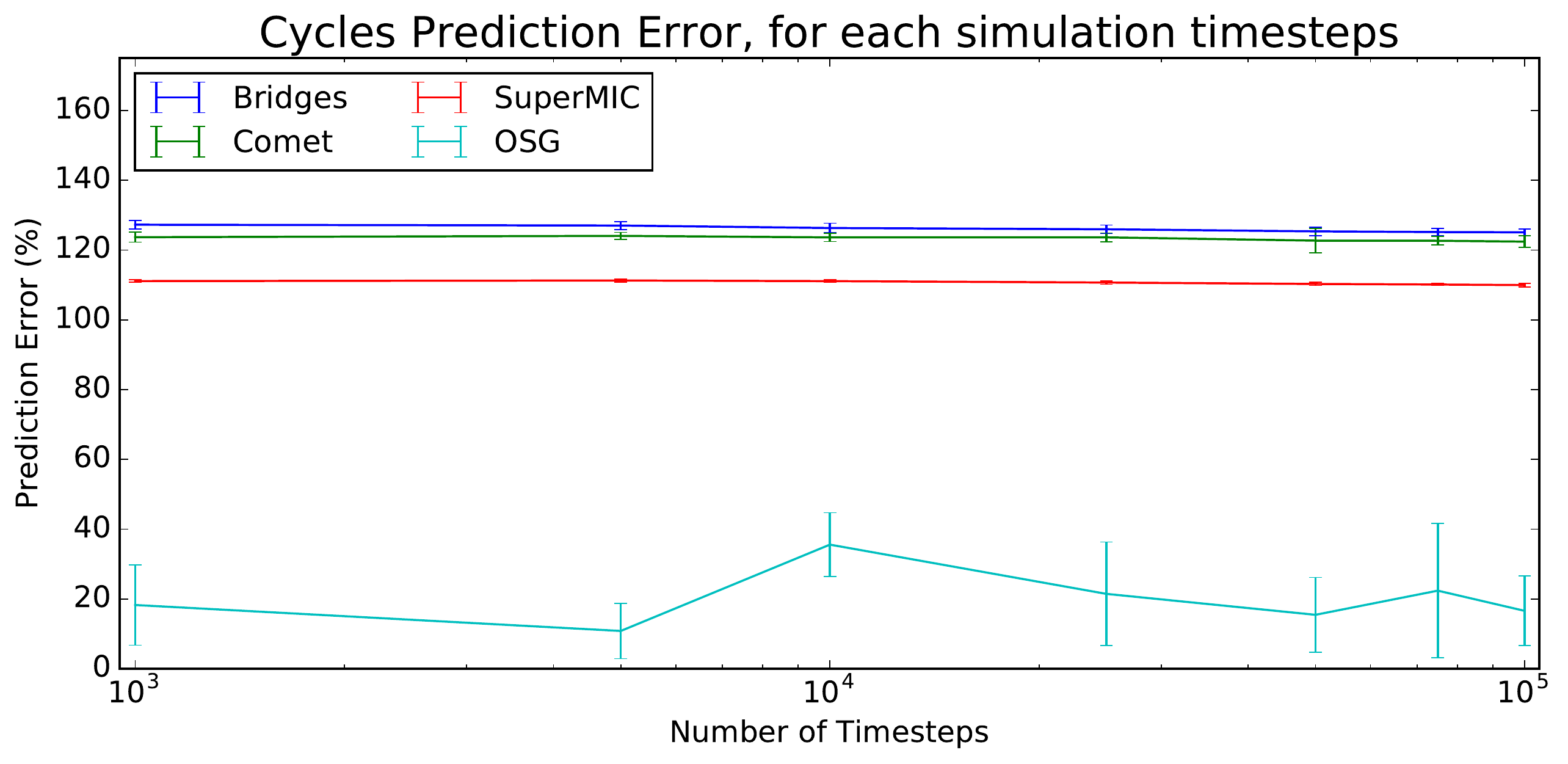}
    \caption{Prediction error of the number of cycles required to run 
    GROMACS simulations on Bridges, Comet, SuperMIC and
    OSG}\label{fig:avg_cycles_err}
\end{figure}

\begin{figure}
    \centering
    \begin{subfigure}[b]{0.47\textwidth}
        \includegraphics[width=\textwidth]{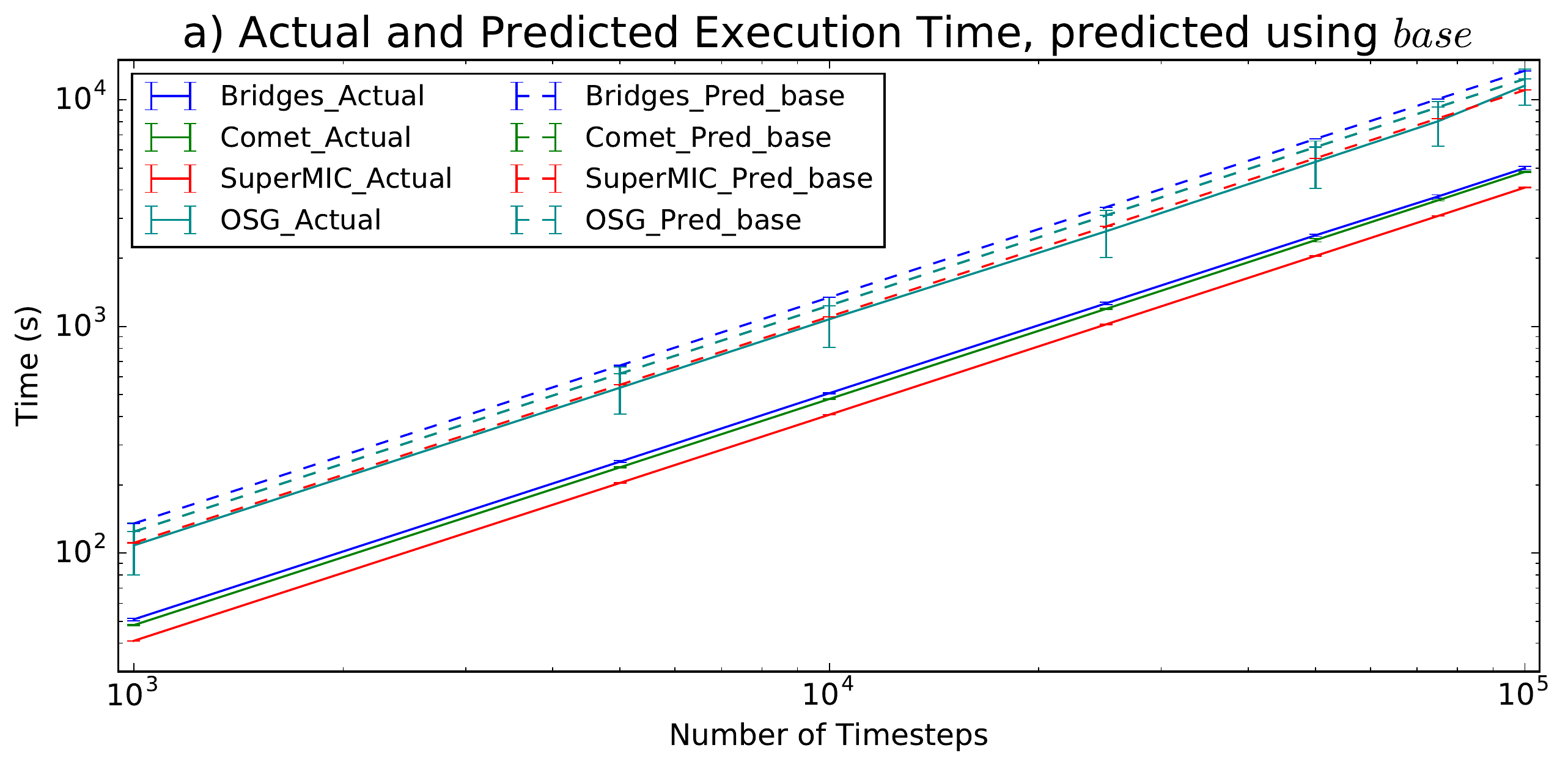}\label{fig:avg_time_pred_base}
    \end{subfigure}
    \begin{subfigure}[b]{0.47\textwidth}
        \includegraphics[width=\textwidth]{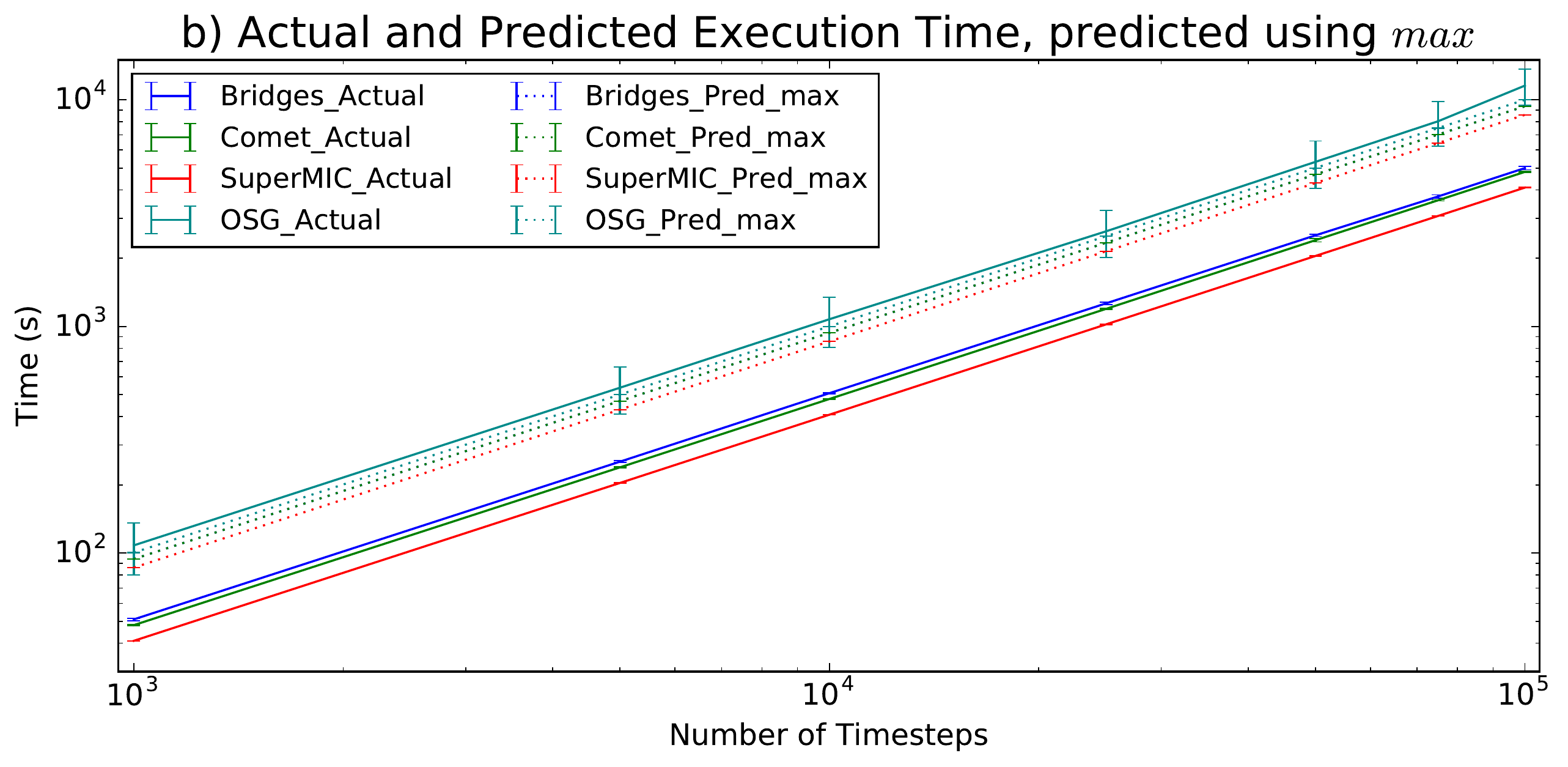}\label{fig:avg_time_pred_max}
    \end{subfigure}
    \caption{Average actual execution time and predicted execution times 
    (using \(base\) and \(max\) frequencies) of GROMACS simulations 
    running on Bridges, Comet, SuperMIC and OSG.}\label{fig:avg_time}
\end{figure}

\begin{figure}
    \centering
    \includegraphics[width=0.47\textwidth]{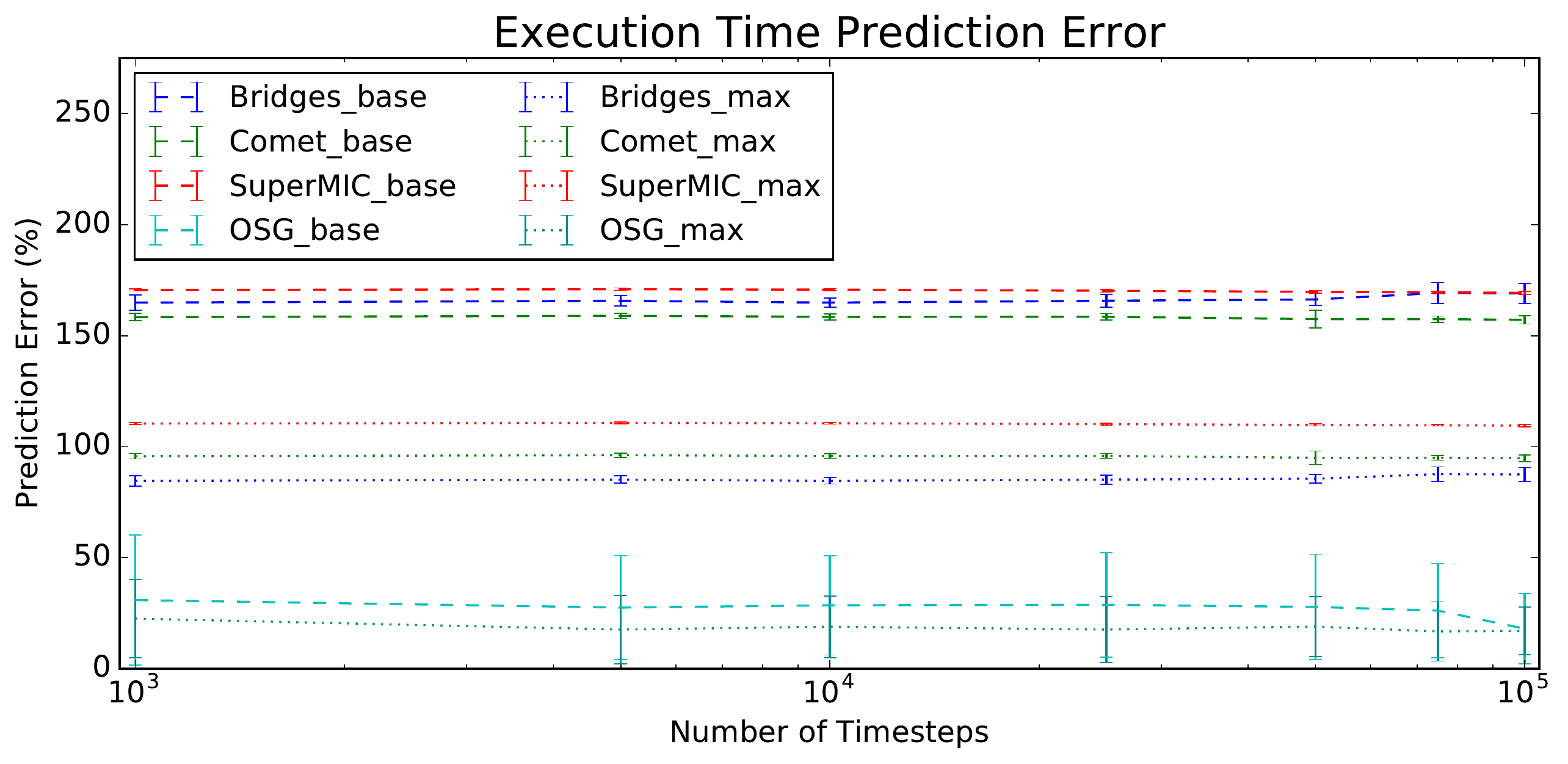}
    \caption{Prediction errors of the predicted execution times (using 
    \(base\) and \(max\) frequencies) of GROMACS simulations running 
    on Bridges, Comet, SuperMIC and OSG}\label{fig:avg_time_err}
\end{figure}

Fig.~\ref{fig:avg_cycles}--\ref{fig:avg_time_err} give a summary of our
findings. All values are shown in the figures as averages, along with their
sample standard deviation as error bars. Fig.~\ref{fig:avg_cycles} shows the
number of cycles required to execute a simulation on Amarel and on the target
machines, as well as the predicted number of cycles required to execute the
simulation sequentially (derived from Amarel data using
Eq.~\ref{eq:p2a_instr_rate}).

Fig.~\ref{fig:avg_instr_rate} shows the instruction rate (i.e., the number of
instructions executed per cycle) for simulations executed on the target
machines, and allows us to predict the number of cycles needed to run the
task on those machines.

Fig.~\ref{fig:avg_cycles_err} shows that our model overpredicts the actual
number of cycles needed to execute the simulations on XSEDE HPC machines by
about 110--125\%. This is unsurprising because our model does not take into
account the code structure of the GROMACS simulation and the hardware
architecture of the the target machines. By calculating \(p2a\_cy\) for each
simulation, we find that the average \(\varepsilon\) for the simulations on
each XSEDE HPC is less than 3\%. This means that our model overpredicts
because we do not consider information that describes the instruction-level
parallelism which the code exploits in the hardware. We also see that our
model provides more reasonable predictions of the number of cycles needed to
execute simulations on OSG\@. However, this is most likely because we
underpredicted the number of instructions required to execute a GROMACS
simulation on OSG\@.

Fig.~\ref{fig:avg_time} and~\ref{fig:avg_time_err} show that our model overpredicts the
task's execution time on XSEDE HPC machines by 157--171\% when using \(base\)
and by 84--111\% when using \(max\). It is important to note that using
\(max\) is more accurate because our model overpredicts the number of cycles
required: Using faster clock frequencies masks the error introduced by
overprediction.

Table~\ref{tab:clkspeed} shows that values of \(avg\) measured on Bridges and
Comet are closer to \(base\) than to \(max\). We also find that our model
overpredicts the task's execution time on OSG by 7--18\% when using \(base\)
and underpredicts by 4--14\% when using \(max\).

Despite our model's tendency to overpredict a task's \(T_x\), we can still
use this model to select resource(s) if for a task \(t\) and any two
resources \(A\) and \(B\), both the actual and predicted \(T_x\) of \(t\)
executed on \(A\) is less than the actual and predicted \(T_x\) of \(t\)
executed on \(B\).

Fig.~\ref{fig:avg_time} shows that our model satisfies the above property
when using \(base\) to predict \(T_x\) on the XSEDE HPC machines, but not
when using \(max\). When using the \(max\) to predict \(T_x\), the predicted
\(T_x\) for both Comet and Bridges are the same because the \(max\) are the
same. However, the actual \(T_x\) measured on Comet is less than that on
Bridges because \(avg\) measured frequency on Comet is less than that on
Bridges.

When using \(base\) to predict \(T_x\) on XSEDE HPC machines and OSG, we find
that our model is inconsistent because the predicted \(T_x\) on OSG is equal
to that on Comet and less than that on Bridges; however, the actual \(T_x\)
on OSG is greater than that on both Comet and Bridges. This is because
\(\sim\)22--24\% more instructions are executed on OSG when running the same
task. To enable a more direct comparison, we increased our prediction on the
number of cycles on OSG by 22\% to account for the additional number of
executed instructions. Doing so makes the \(T_x\) predictions generated using
\(base\) consistent across HPC machines and OSG\@.

% ---------------------------------------------------------------------------
\subsection{Resource Selection}

We performed a set of experiments comparing the cost of performing resource
selection on the basis of our model's predictions to the cost of a random
resource selection. We expressed the cost of resource selection in terms of
time-to-completion of the workload and we executed a bag-of-task workload on
XSEDE HPC machines (Bridges, Comet, SuperMIC) and on OSG\@. Again, we call
the XSEDE HPC machines and OSG as the target machines. Each task of the
workload consisted of a GROMACS simulation running for \(10^5\) timesteps as
specified in Sec.~\ref{sec:tx_char}.

The metric of performance used by the model is the time-to-completion of the
task \(TTC_{task}\): resources selected on the base of our model have the
smallest predicted \(TTC_{task}\). We define \(TTC_{task} = T_{q,task} +
T_{x,task}\), where \(T_{q,task}\) and \(T_{x,task}\) are the time taken to
acquire the resources necessary to execute a task and the time taken to
execute the task on the acquired resources, respectively.

Though any measure of cost can be used to decide how to select resources,
workload's time-to-completion is one of particular interest to users. We
define the time-to-completion of a workload \(TTC_{wkd}\) as \(T_{q, wkd} +
T_{x, wkd}\), where \(T_{q, wkd, act}\) is the amount of time spent acquiring
resources and \(T_{x, wkd}\) is the amount of time spent executing at least
one task. Using \(TTC_{wkd}\) allows to measure how selecting resources that
minimize the time-to-completion of each task affects the time-to-completion
of the entire workload. Since time components of a task's execution time are
\(T_{q,task}\) and \(T_{x,task}\) and no data movement or preprocessing and
post processing occurs, \(TTC_{wkd}\) only involves \(T_{q,wkd}\) and
\(T_{x,wkd}\).

% ------------
\subsubsection{Experimental Setup}

We executed workloads with 64, 128, 256, 512, and 1024 tasks across the
target machines, repeatedly over the course of a month. Only one distributed
execution of the workload occurred at any given time, preventing
self-competition for resource acquisition. We used RADICAL-Pilot
(RP)~\cite{rp} to concurrently acquire resources across the target machines
and to distribute the execution of the workload across those resources.

We submitted at most one pilot~\cite{pilot-abstr} to the local resource
manager of any XSEDE HPC machine to acquire the resources (e.g., cores and
walltime) necessary to execute the entire workload. Concurrently submitting
multiple pilots to the same HPC machine would have created self-competition
for resources, requiring further investigation of the effects of pilot sizing
on the distributed execution of the workload~\cite{jobsizing}. On OSG, we
submitted only single-core pilots, enough to acquire the number of cores
required to execute the entire workload. While it is possible to submit
multi-core pilots to OSG, the XSEDE OSG documentation recommends against it.

When acquiring cores from XSEDE HPC machines, their resource managers return
the smallest number of nodes with the number of cores requested. If the
number of tasks assigned to run on an HPC machine is not a multiple of the
number of cores per node, some tasks execute on a node with unused cores. We
found that on SuperMIC, the \(T_x\) of our GROMACS simulations varies up to
\(\sim\)(19\%), depending on whether all the cores of a node are utilized.
To control this fluctuation in \(T_x\) we used additional ``padding tasks''
to occupy all the cores of each compute node.

We used our cost model of task execution to predict \(T_{x,task}\) and
therefore derive \(TTC_{task}\). For the values of \(T_{x,task}\), we used
the predictions of the execution time of a simulation running on the
resources of the target machines derived using \(base\) frequencies. Note
that we found in Sec.~\ref{sec:tx_char} that a simulation running a fixed number
of timesteps performs \(\sim\)22--24\% more instructions when executed on OSG
than when executed on XSEDE HPC machines. Accordingly, we increased the
predicted number of cycles required to execute a simulation on OSG by
\(\sim\)20\%. 

We used XDMoD~\cite{xdmod} to collect historical information about queue
waiting times of jobs submitted to the XSEDE HPC machines to derive values
for \(T_{q,task}\). Since XDMoD does not provide any historical data on the
queue waiting times of jobs submitted to OSG, we used a sample of trial runs
for jobs submitted to OSG to calculate \(T_{q,task}\). XDMoD allows us to
filter historical data of queue waiting times of jobs based on the queue and
machine to which the jobs were submitted, as well as the walltime and the
number of cores requested.

We calculated values for the \(T_{q,task}\) of a task as the average queue
waiting time of jobs submitted to the same queue of the same machine that
requested a `similar' walltime and number of cores within the past 7 days.
When filtering data on XDMoD based on requested job walltime or requested
number of cores, XDMoD automatically clusters the data points into predefined
ranges and limits the granularity with which we filter data. Thus, we
consider two jobs to have similar requested job walltime or requested number
of cores when the values of the requested job walltime (or number of cores)
fall within the same predefined range. We find that for jobs requesting a
large number of cores (e.g. 512, 1024 cores), there is often missing data
because no jobs of that size were submitted. In this case, we used data
points from jobs that were submitted to the same machine and queue.

% ------------
\subsubsection{Results}

\begin{figure}
    \begin{subfigure}[b]{0.5\textwidth}
        \includegraphics[width=\textwidth]{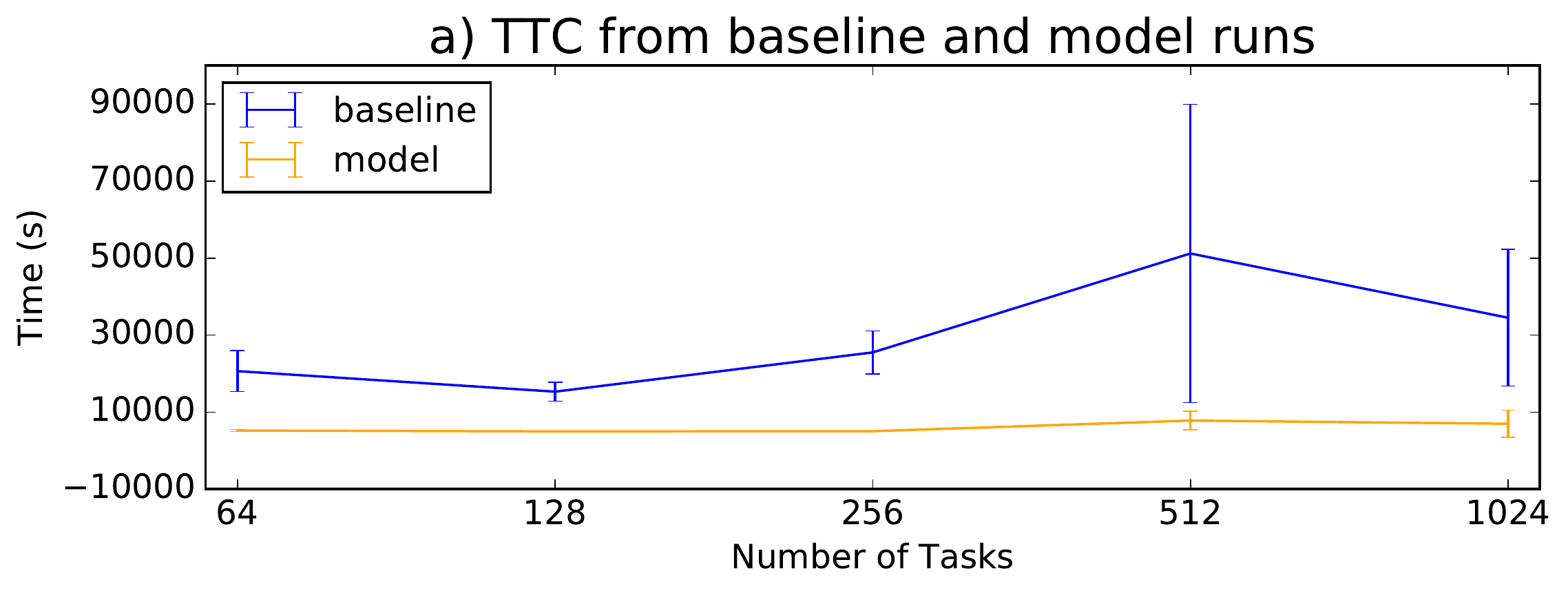}
        \caption{\(TTC_{wkd}\)}\label{fig:avg_ttc_wkd}
    \end{subfigure}
    \begin{subfigure}[b]{0.5\textwidth}
        \includegraphics[width=\textwidth]{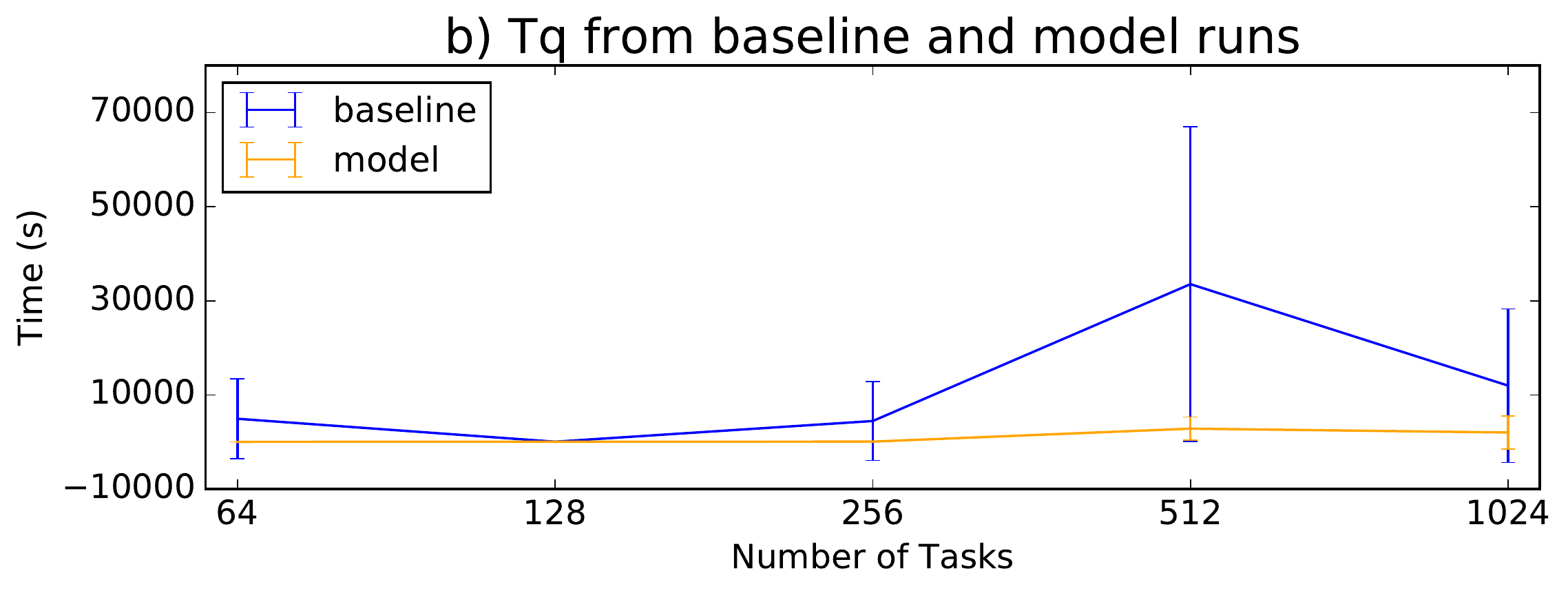}
        \caption{\(T_{q,wkd}\)}\label{fig:avg_ttq_wkd}
    \end{subfigure}
    \begin{subfigure}[b]{0.5\textwidth}
        \includegraphics[width=\textwidth]{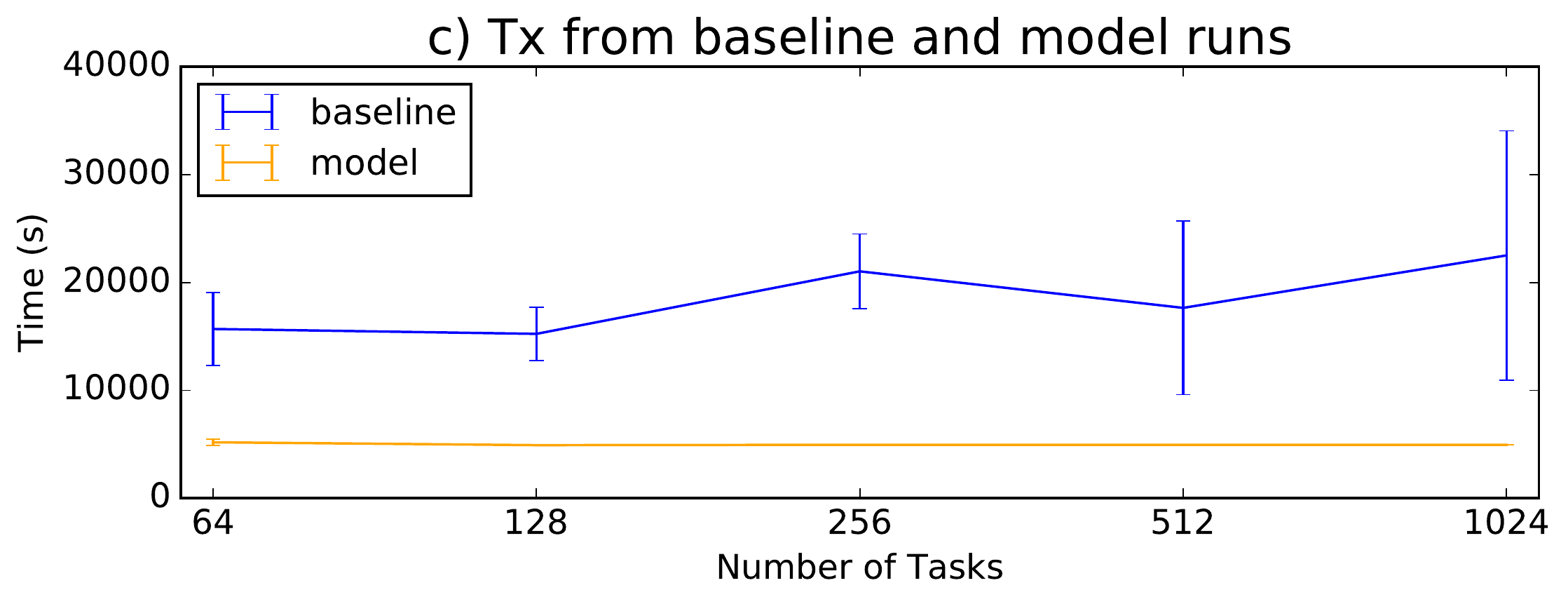}
        \caption{\(T_{x,wkd}\)}\label{fig:avg_ttx_wkd}
    \end{subfigure}
    \caption{\(TTC_{wkd}\), \(T_{q, wkd}\) and \(T_{x, wkd}\) measured from 
    the baseline runs and model runs.}\label{fig:avg_res_select}
\end{figure}

Fig.~\ref{fig:avg_res_select} shows the average \(TTC_{wkd}\), \(T_{q, wkd}\)
and \(T_{x, wkd}\) of the runs we performed over a month. We call the set of
runs where the workload was distributed randomly the `baseline' runs, and the
set of runs where the workload was distributed using the cost model the
`model' runs. The error bars denote the sample standard deviation. During the
period in which we executed the workload using our model, the target machine
selected by our model using \(TTC_{task}\) was SuperMIC because values of
\(T_{q,task}\) and \(T_{w,task}\) were consistently lower than those of all
the other target machines.

Fig.~\ref{fig:avg_ttc_wkd} shows that the average \(TTC_{wkd}\) measured when
distributing a workload across target machine(s) selected either randomly or
using our cost model. From Fig.~\ref{fig:avg_ttc_wkd}, we see that the
average \(TTC_{wkd}\) of executing a workload when the workload was
distributed using our model is 67--85\% lower than when executing the same
workload when the workload was distributed randomly. It is important to note
that the values of \(T_{x,task}\) from the predictions of \(T_x\) were
generated using the \(base\) frequencies of the target machines. These
predictions overpredicted the actual \(T_x\) by 157--171\% (shown in
Fig.~~\ref{fig:avg_time_err}). This shows that even inaccurate predictions of
task execution time can be consistently used to select resources that support
a lower workload time-to-completion than that obtained with a random resource
selection.

Fig.~\ref{fig:avg_ttq_wkd} shows the average \(T_{q,wkd}\) of the baseline
and model runs. From Fig.~\ref{fig:avg_ttq_wkd}, we see that the average
\(T_{q,wkd}\) and the sample standard deviation measured by the models runs
are smaller than those measured by the baseline runs. This is due to the
different queue waiting time across machines and the delay it introduces for
task execution. Tasks randomly distributed across multiple machines waiting
longer to execute than tasks submitted only to SuperMIC on the base of our
model's predictions. Consistently with other predictions and the historical
data of XDMoD, during the month of our experimental campaign, SuperMIC's
queue waiting time were on average much lower and more stable than that
measured on Bridges and Comet. This explains the comparatively smaller
average \(T_{q,wkd}\) and standard deviation of the model runs.

Note that the \(T_{q,wkd}\) measured from the baseline runs is sensitive the
the size of the requested resources. Queue waiting time for runs with 512
tasks/cores was consistently longer on Comet and Bridges than on SuperMIC\@.
Further, runs with 128 tasks/cores had consistently much lower queue waiting
time on all the three HPC machines. These differences may account for the
variations of the sample standard deviation of \(T_{q,wkd}\) in the baseline
runs. Additional experiments are required to confirm the relationship between
the sample standard deviation of \(T_{q,wkd}\) and workload size.

Fig.~\ref{fig:avg_ttx_wkd} shows average \(T_{x,wkd}\) and sample standard
deviations measured by the baseline and model runs. For the model runs, we
find that the values of \(T_{x,wkd}\) for each workload size are almost
identical, and that the sample standard deviation is negligible. Again, this
is because all tasks were assigned to execute on SuperMIC\@. We find that the
\(T_{x,wkd}\) and larger sample standard deviations are higher measured from
the baseline runs are higher than those measured from the model runs. One
reason for this is that workloads that are distributed randomly assigned
tasks to run on OSG. Tasks running on OSG have a larger execution time than
when they run on any other target machine. Since OSG is a collection of
resources, the execution time of the task running on an OSG resource can
vary.

The second reason is that \(T_{x, wkd}\) measures the amount of time where at
least one task of the workload is executing. We find that the queue waiting
time of pilots are staggered in time. For the baseline runs, it is common to
find little or no time overlap between the execution of tasks that have been
assigned to run on different DCRs. For the model runs, we make submit only
one pilot to SuperMIC to acquire resources to execute the entire workload.
When the pilot becomes active, all tasks in the workload execute
concurrently. It is important to note this is true because we are considering
homogeneous tasks. When considering homogeneous tasks, assigned all tasks to
execute on one machine.

%% file: conclusions.tex
In this paper, we present a model that can predict the cost of executing a
task on a resource. The model does not require information like code
structure or hardware architecture. Although this limits the accuracy of its
predictions, it enables the prediction of task execution times on
\textit{new} target resources using historical data collected on a baseline
resource. Currently, the model does not take into account any parallelism
which a task exploits during its execution. Extending the model to represent
tasks using multithreading or multiprocessing is considered future work.

Experimental results show that the model consistently over-predicts the
actual execution time of a GROMACS MD simulation running one thread and
process by 157--171\%. Nonetheless, the predictions can still be used to
determine which resource yields the smallest execution time.

We incorporated the model into the Condor matchmaking algorithm to address
the resource selection problem. The Condor matchmaking algorithm is used
primarily in resource brokers to determine whether a job can run on a given
resource. By adapted the matchmaking algorithm with our model, we can use the
algorithm to determine which resource to use to execute a task.

We used task execution cost model and the adapted matchmaking algorithm to
distribute a bag-of-task workload of GROMACS MD simulations across both HPC
and HTC resources based on the expected time-to-completion of each task of
the workload. For workloads of up to 1024 GROMACS simulations, our results
show a reduction in the time-to-completion by 67--85\% compared to randomly
distributing the workload across the same resources. This shows that
inaccurate predictions of execution times can still be used to select
resources better than random. Moreover, it is possible to select resources
where we have no historical data better than random.

Our results demonstrate the usefulness of our approach on XSEDE, but they are
not limited to traditional distributed resource. For example, our resource
selection models could be used to select heterogeneous virtual machine
``instances'' from federated cloud resources and metrics such as (fiscal)
costs of instances. These extensions will be useful as the WLCG moves to
incorportate cloud resources and spot markets into their mix of resources
(HEPCloud~\cite{hepcloud}).

%% file: acknowledgment.tex
\footnotesize{\textit{Acknowledgements:} MTH is supported by a GAANN graduate
fellowship. This work is also funded by NSF CAREER ACI-1253644 and Department
of Energy Award ASCR DE-SC0016280. Computational resources were provided by
NSF XRAC award TG-MCB090174.